\documentclass[reprint, twocolumn, superscriptaddress,prb]{revtex4-2}
\usepackage{bm, amsmath, amsfonts, amssymb, braket}
\usepackage{subfigure}
\usepackage{color}
\usepackage{graphicx}
\usepackage{slashed}

\usepackage{dcolumn} % Align table columns on decimal point

\usepackage{rotating} % for 90 degree rotated table and fig

\usepackage{ulem}
\usepackage{xcolor}
\DeclareRobustCommand{\erase}{\bgroup\markoverwith{\textcolor{red}{\rule[.5ex]{2pt}{0.4pt}}}\ULon}

\begin{document}
%\title{Optimize quantum geometry of Chern insulator by lattice reconfiguration}
\title{Biorthogonal basis approach to fractional Chern physics}
\author{Nobuyuki Okuma}
\email{okuma@hosi.phys.s.u-tokyo.ac.jp}
%\author{Masatoshi Sato}
%\altaffiliation{Department of Physics, University of Tokyo, Hongo 7-3-1, 113-0033, Japan}%Lines break automatically or can be forced with \\
\affiliation{%
  Graduate School of Engineering, Kyushu Institute of Technology, Kitakyushu 804-8550, Japan
 %This line break forced with \textbackslash\textbackslash
}%

\date{\today}
\begin{abstract}
A fractional Chern insulator is thought to emerge from the competition between one-particle band topology and strong repulsive interactions. As an attempt to study lattice models of fractional Chern insulators, we introduce a biorthogonal basis constructed from coherent-like states on the von Neumann lattice. Focusing on the fact that this basis is diagonal to the vortex attachment in the infinite-volume limit, we convert the original fermions into composite fermions by applying a two-dimensional Jordan-Wigner transformation to the creation and annihilation operators of the biorthogonal basis. Furthermore, we apply the Hartree-Fock mean-field approximation to handle the interaction Hamiltonian of composite fermions. Due to the biorthogonal nature, the representation of the new Hamiltonian is no longer Hermitian, which implies that the introduction of the approximation does not guarantee the reality of the energy spectrum. In fact, there are many self-consistent solutions with complex spectra. Nevertheless, we numerically find that it is possible to construct a self-consistent solution where the band dispersion is nearly real and the ground-state energy is lower than in other solutions.

\end{abstract}
\maketitle
\section{Introduction}
The study of topological phases has gained significant attention in recent years. Among the various phenomena, the integer \cite{Klitzing-80} and fractional quantum Hall effects (IQHE and FQHE) \cite{FQHE-exp-82,Laughlin-Wavefunction-83,yoshioka-textbook-02,halperin-fractional-textbook-20,fradkin2013field} are regarded as foundational and pioneering examples.
From a theoretical perspective, these phenomena benefit from band-flatness and analytical properties of Landau levels, which have led to extensive research on them from the time of their discovery to the present day.
Recently, these concepts have been extended to topological insulators \cite{Kane-review,Zhang-review} and intrinsic topological order \cite{wen1990topological}.
While the former can be explained within the one-particle picture, the latter is essentially understood through many-body physics.

Among the topological insulators, the Chern insulator~\cite{Haldane-88} is the simplest example in two dimensions and can be considered topologically equivalent to the IQHE.
While the nontrivial topology in the IQHE comes from a strong magnetic field, that of the Chern insulator originates from the band topology and does not need the magnetic field.
Focusing on the fact that the difference between the IQHE and FQHE lies in whether the filling factor $\nu$ is an integer or a fraction, one might expect that a fractionally-filled Chern insulator is topologically equivalent to the FQHE.
A material in which an FQHE-like state is realized in the absence of the magnetic field is called a fractional Chern insulator (FCI) \cite{Regnault-Bernevig-11, Bergholtz-Liu-13, Parameswaran-13, Liu-Bergholtz-review-22}.
Although a fully established FCI has not yet been experimentally realized, very recent reports have shown promising experimental results in two-dimensional materials \cite{li2021spontaneous,cai2023signatures,zeng2023thermodynamic,park2023observation,xu2023observation,lu2023fractional}.
However, this does not mean all fractionally-filled Chern bands are FCIs.
Generally, Chern bands are completely different from Landau levels, aside from their topological aspects.
For example, Chern bands typically lack the flatness and analytic properties that exist in Landau levels.
Searching for materials that exhibit such properties at the non-interacting level is the first step in the search of the FCI, and much research has been conducted in this area \cite{Regnault-Bernevig-11, Bergholtz-Liu-13, Parameswaran-13, Liu-Bergholtz-review-22,Parameswaran-Roy-Sondhi-12,Roy-geometry-14,Jackson-Moller-Roy-15,Claassen-Lee-Thomale-Qi-Devereaux-15,Lee-Claassen-Thomale-17,Mera-Ozawa-21-2,Varjas-Abouelkomsan-Kang-Bergholtz-22}.
Among them, the notion of ``vortexability" \cite{ledwith2023vortexability} provides one of the simplest pictures for describing the distance between a given Chern band and Landau levels.
It is well known that the operation of multiplying a state by a complex coordinate $z:=x+iy$ remains closed within the LLL.
Similarly, the operation of multiplying a state by a ``vortex function", which is the generalization of the complex coordinate, is closed within the ``vortexable" Chern band. This notion can also be defined for the higher Landau levels \cite{fujimoto2024higher}.
Since the physics of the FQHE has a history of being discussed in terms of complex coordinates, describing FCIs using vortex functions is a reasonable approach.

In our previous work \cite{okuma2024constructing}, we generalized the notion of the vortex function to the lattice models, where the function (or diagonal operator in real space) becomes a diagonal matrix on the position basis.
We called it the lattice vortex function.
Moreover, we introduced the notion of coherent-like states \cite{okuma2024constructing} on von Neumann lattice \cite{von2018mathematical,perelomov2002completeness, bargmann1971completeness}, defined by the set of (left) eigenstates of the projected lattice vortex function.
These states are the direct generalization of the true coherent states that are the eigenstates of the annihilation operator in the Landau levels \cite{imai1990field,ishikawa1992field,ishikawa1999field}.
Remarkably, the difference between the problem of Landau levels in continuous systems and that of Chern bands in lattice systems is reduced to the distinction between coherent states and coherent-like states, both of which are defined on the same lattice.
This formulation may obscure the analytic properties, but it holds the potential to provide a unified description of the FQHE and FCI.

In this paper, we discuss the composite fermionization on a biorthogonal basis constructed from the coherent-like states.
We first define the biorthogonal basis as the set of eigenstates of the vortex function that is projected onto the Chern band.
Since this basis is diagonal in the vortex attachment, one can expect that composite fermions \cite{jain2007composite} can naturally be introduced.
For the composite fermionization, we apply a two-dimensional Jordan-Wigner transformation \cite{fradkin1989jordan} to the fermionic operators that create and annihilate the biorthogonal-basis states on the von Neumann lattice.
Using this composite fermionization and mean-field approximation, we investigate several $C=1$ Chern-insulator models with the filling factor $\nu=1/3$, under a short-range repulsive interaction.
Ideally, the composite-fermionic Hamiltonian also becomes a $C=1$ Chern insulator.
Because of the biorthogonal nature, the representation of the effective quadratic Hamiltonian is no longer Hermitian. Under the non-Hermiticity, there are a lot of self-consistent solutions whose spectra are not even real.
Nevertheless, we find that it is possible to construct a self-consistent solution where the energy is almost real and the ground-state energy is lower than in other solutions.

This paper is organized as follows.
In Sec.\ref{conventionsection}, we introduce several conventions and notations used in this paper.
In Sec.\ref{biorthogonalsection}, we develop a theory of biorthogonal basis on the von Neumann lattice of the Chern insulator.
The purpose of this section is to present part of our previous work while also addressing the finer points that were not fully resolved in that paper.
In Sec.\ref{applicationsection}, we introduce the composite fermionization on the biorthogonal basis and apply it to the fractional Chern physics.

\section{Convention and notation\label{conventionsection}}
As a kinetic part, we consider a fermionic quadratic lattice model given by:
\begin{align}
    \hat{H}=\sum_{\bm{R},\bm{R}'}\sum_{i,j}c^\dagger_{\bm{R},i}H_{(\bm{R},i),(\bm{R}',j)} c_{\bm{R}',j}, \label{tight}
\end{align}
where $\bm{R}=(X,Y)$ and $i,j=1,2,\cdots,n_{\rm orb}$ denote the unit cell vector and intracell atomic orbital, respectively.
Since our motivation is on solid-state physics, we impose discrete translation symmetry on Eq. (\ref{tight}).
As is well known, there are two common conventions for the Fourier transform:
\begin{align}
    &c^{\dagger}_{\bm{k},i}=\frac{1}{\sqrt{N_{\rm unit}}}\sum_{\bm{R}}e^{i\bm{k}\cdot\bm{R}}c^{\dagger}_{\bm{R},i},\label{indep}\\
    &\tilde{c}^{\dagger}_{\bm{k},i}=\frac{1}{\sqrt{N_{\rm unit}}}\sum_{\bm{R}}e^{i\bm{k}\cdot\bm{R}+\bm{r}_i}c^{\dagger}_{\bm{R},i},\label{position-dependent}
\end{align}
where $\bm{k}$ is the crystal momentum, $N_{\rm unit}$ is the number of unit cells, and $\bm{r}_i=(x_i,y_i)$ is the intracell sublattice position of the orbital $i$. We adopt the convention (\ref{indep}). Under the discrete translation symmetry, the Hamiltonian (\ref{tight}) can be rewritten as
\begin{align}
\hat{H}=\sum_{\bm{k}}\sum_{i,j}c^\dagger_{\bm{k},i}[H_{\bm{k}}]_{i,j}
    c_{\bm{k},j}.
\end{align}
The one-particle dispersion is obtained by diagonalizing the Bloch Hamiltonian matrix $H_{\bm{k}}$:
\begin{align}
    H_{\bm{k}}=\sum_{\alpha}\epsilon_{\bm{k},\alpha}\ket{u_{\bm{k},\alpha}}\bra{u_{\bm{k},\alpha}},
\end{align}
where $\alpha$ denotes the band index, and $\ket{u_{\bm{k},\alpha}}$ are the periodic parts of the Bloch eigenstates.
In the position basis $\{\ket{\bm{R},i}\}$, the Bloch eigenstates $\{\ket{\bm{k},\alpha}\}$ are represented as
\begin{align}
    \langle \bm{R},i\ket{\bm{k},\alpha}=u_{\bm{k},\alpha}(i)\frac{e^{i\bm{k}\cdot\bm{R}}}{\sqrt{N_{\rm unit}}},
\end{align}
where $u_{\bm{k},\alpha}(i):=\langle i\ket{u_{\bm{k},\alpha}}$. 
We assume that $\ket{u_{\bm{k},\alpha}}$ and $\ket{\bm{k},\alpha}$ are normalized.

In addition to the Fourier transform, there is also a convention issue regarding the sign of the Berry connection in band theory.
We here define the Berry connection, Berry curvature, and Chern number as
\begin{align}
    a_I(\bm{k},\alpha)&=-i\bra{u_{\bm{k}},\alpha}\partial_{k_{I}}\ket{u_{\bm{k}},\alpha},\\
    \omega(\bm{k},\alpha)&=\frac{\partial a_y(\bm{k},\alpha)}{\partial k_x}-\frac{\partial a_x(\bm{k},\alpha)}{\partial k_y},\\
    C_{\alpha}&=\int_{\rm BZ} \frac{d^2k}{2\pi}\omega(\bm{k},\alpha),
\end{align}
where $I=x,y$, and BZ denotes the Brillouin zone.
Note that the Berry connection and curvature depend on the choice of the Fourier convention:
\begin{align}
    A_I(\bm{k},\alpha)&=-i\bra{U_{\bm{k}},\alpha}\partial_{k_{I}}\ket{U_{\bm{k}},\alpha}\notag\\
    &=a_I(\bm{k},\alpha)-\sum_{i}r^I_iu^*_{\bm{k}}(i)u_{\bm{k}}(i),\label{posi-berry}\\
    \Omega(\bm{k},\alpha)&=\frac{\partial A_y(\bm{k},\alpha)}{\partial k_x}-\frac{\partial A_x(\bm{k},\alpha)}{\partial k_y},
\end{align}
where $\ket{U_{\bm{k},\alpha}}$ is the periodic part of a Bloch eigenstate in the convention (\ref{position-dependent}).
Nevertheless, topological numbers such as the Chern number, which are globally defined, do not rely on the convention.

As an interacting part, we consider the following repulsive Hamiltonian:
\begin{align}
    H_{\rm int}&=\frac{1}{2}\sum_{i,j,\bm{r},\bm{r}'}V^{ij}(\bm{r}-\bm{r}')\rho_{\bm{r},i}\rho_{\bm{r}',j}\notag\\
    &=\frac{1}{2N_{\rm unit}}\sum_{i,j}\sum_{\bm{q}}V^{ij}(-\bm{q})\rho_{\bm{q},i}\rho_{-\bm{q},j}.
\end{align}
The density operators are defined as
\begin{align}
    &\rho_{\bm{r},i}=c^{\dagger}_{\bm{r},i}c_{\bm{r},i}=\frac{1}{N_{\rm unit}}\sum_{\bm{q}}e^{i\bm{q}\cdot\bm{r}}\rho_{\bm{q},i},\\
    &\rho_{\bm{q},i}=\sum_{\bm{r}}e^{-i\bm{q}\cdot\bm{r}}\rho_{\bm{r},i},\label{densityoperator}\\
    &V^{ij}(-\bm{q})=\sum_{\Delta\bm{r}}e^{i\bm{q}\cdot(\Delta\bm{r})}V^{ij}(\Delta \bm{r}),
\end{align}
where $\Delta\bm{r}:=\bm{r}-\bm{r}'$. 

In our paper, we consider some non-Hermitian matrices (operators).
Under the non-Hermiticity, the Hermitian conjugate of the ket eigenvector is not always a bra eigenvector.
Therefore, we define the right and left eigenvectors as
\begin{align}
    &M\ket{\lambda,R}=\lambda\ket{\lambda,R},\notag\\
    &\bra{\lambda,L}M=\lambda\bra{\lambda,L},
\end{align}
where $M$ is a matrix or an operator.
If we use the word ``mode" instead of the eigenstate, it means a ``right" eigenstate.
Note that for $\lambda$ in the spectrum of $M$, the existence of the left or right eigenvectors is not always guaranteed.
The exact zero modes discussed in Sec.\ref{zeromodesection} are a typical example of such a situation.

\section{Biorthogonal basis on von Neumann lattice\label{biorthogonalsection}}
In this section, we first introduce the exact zero mode of (hermitian conjugate of) projected vortex function whose origin is topology. By using the exact zero mode, we construct the biorthogonal basis as an analogy of the von Neumann lattice basis.
This section is a reorganization of a part of our previous work \cite{okuma2024constructing} with some additional elements.

\subsection{Lattice vortex function}
In the position basis $\{\ket{\bm{R},i}\}$, the lattice vortex function is represented as \cite{okuma2024constructing}
\begin{align}
    Z&=\alpha_1~ (X+\sum_i\tilde{x}_iP_i)+\alpha_2~ (Y+\sum_i\tilde{y}_iP_i)\notag\\
    &=\alpha_1~ X+\alpha_2~ Y +\sum_i\gamma_iP_i,\label{vortexfunction}
\end{align}
where $\alpha_1,\alpha_2,\gamma_i\in\mathbb{C}$ are parameters, and $P_i:=\ket{i}\bra{i}$ is the projection operator onto sublattice $i$. 
Physically, $\tilde{\bm{r}}_i=(\tilde{x}_i,\tilde{y}_i)$ is a virtual sublattice position, which can be different from the real sublattice position.
The lattice vortex function is a direct analogy of the vortex function introduced for continuous models \cite{ledwith2023vortexability}.
In our lattice case, $Z$ is a diagonal matrix rather than a function.
If we set $(\alpha_1,\alpha_2)=(1,i)$ and $\tilde{\bm{r}}_i=\bm{r}_i$, Eq. (\ref{vortexfunction}) becomes a lattice analogy of the the complex coordinate $z=x+iy$ in the QHE context.
The parameters of the lattice vortex function $\alpha_1,\alpha_2,\gamma_i\in\mathbb{C}$ should be determined based on some criteria, such as ideal conditions \cite{Roy-geometry-14,Jackson-Moller-Roy-15} and vortexability \cite{ledwith2023vortexability}.
If the determined $\tilde{\bm{r}}_i$ is different from the original position $\bm{r}_i$, the difference describes the non-uniformity in space metric, which is discussed in terms of $r$-ideal Chern band in continuous models \cite{estienne2023ideal}.

\subsection{Projected vortex function}
In the following, we focus on the physics projected onto one Chern band $\alpha$ with the Chern number $C>0$, while the extension to cases where the number of topological bands exceeds one is trivial \cite{okuma2024constructing}.
The projection operator onto the Chern band is given by
\begin{align}
    P=\sum_{\bm{k}}\ket{\bm{k}}\bra{\bm{k}}.
\end{align}
For the sake of simple descriptions, we will omit the band index $\alpha$ henceforth.
We first consider the infinite-volume limit with no boundary.
The wave functions on the Chern band are given by the following wavepacket form:
\begin{align}
    \ket{\psi}\propto\int\frac{d^2k}{(2\pi)^2}a(\bm{k})\ket{\bm{k}}.
\end{align}
In other words, the coefficient $a(\bm{k})$ is the representation of the wave function in the crystal-momentum basis.
In momentum space, the position operators (for virtual configuration) are defined as the following differential operators:
\begin{align}
    P(X+\sum_i\tilde{x}_iP_i)P&\leftrightarrow \hat{x}_{\bm{k}}=i\partial_{k_x}-A_x(\bm{k}),\notag\\
    P(Y+\sum_i\tilde{y}_iP_i)P&\leftrightarrow \hat{y}_{\bm{k}}=i\partial_{k_y}-A_y(\bm{k}),
\end{align}
where the Berry connection is calculated by replacing $\bm{r}_i$ in Eq. (\ref{posi-berry}) with $\tilde{\bm{r}}_i$.
Thus, the lattice vortex function projected onto the Chern band is also given by the differential operators:
\begin{align}
    PZP&\leftrightarrow \hat{z}_{\bm{k}}=\alpha_{1} \hat{x}_{\bm{k}}+\alpha_2 \hat{y}_{\bm{k}},\\
    PZ^*P&\leftrightarrow \hat{z}^{\dagger}_{\bm{k}}=\alpha^*_{1} \hat{x}_{\bm{k}}+\alpha^*_2 \hat{y}_{\bm{k}}.
\end{align}

\subsection{Exact zero mode of projected vortex function\label{zeromodesection}}
For the construction of the biorthogonal basis, the following exact zero mode(s) are essentially important.
As shown in our previous work \cite{okuma2024constructing}, $PZ^*P$ has at least $C$ exact zero modes in the projected space. 
Here, the word ``mode" means a right eigenstate (see Sec.\ref{conventionsection}).
Let $\nu_{Z^*}$ and $\nu_{Z}$ be the numbers of exact zero modes of $PZ^*P$ and $PZP$ in the projected space, respectively.
According to our previous work \cite{okuma2024constructing}, the following holds:
\begin{align}
    \nu_{Z^*}-\nu_{Z}=C.
\end{align}
In the proof, we have used the index theorem for a Dirac operator on the momentum-space torus:
\begin{align}
i\slashed{D}(\bm{k})=
    \begin{pmatrix}
        0&\hat{z}_{\bm{k}}\\
        \hat{z}^{\dagger}_{\bm{k}}&0
    \end{pmatrix}.
\end{align}
Here, the gauge field in the covariant derivative is nothing but the Berry connection in momentum space.

For simplicity, we assume $C=1$ henceforth.
In this case, $PZ^*P$ has at least one exact zero mode.
Except in the case where $PZP$ has accidental zero modes, $PZ^*P$ has one exact zero mode. In momentum-space language, the coefficient of the zero-mode wavepacket, $a_0(\bm{k})$, satisfies
\begin{align}
\hat{z}^{\dagger}_{\bm{k}}a_0(\bm{k})=0.
\end{align}
Unlike the Wannier states in topological bands, the wavepacket for $a_0(\bm{k})$ is exponentially localized in real space.

In an actual calculation of $a_0(\bm{k})$, it is useful to consider the eigenvalues problem of the following Hermitian operators, rather than the unstable non-Hermitian problem:
\begin{align}
    PZPZ^*P~\mathrm{or}~\hat{z}_{\bm{k}}\hat{z}^{\dagger}_{\bm{k}}.
\end{align}
Needless to say, $PZ^*P$ and $PZPZ^*P$ share the same zero mode.
The non-zero eigenmodes of $PZPZ^*P$ are also discussed in terms of the radially localized basis in our previous work \cite{okuma2024constructing}.
In the case with $(\alpha,\beta)=(1,i)$, $\hat{z}_{\bm{k}}\hat{z}^{\dagger}_{\bm{k}}$ becomes
\begin{align}
    \left[(-i\partial_{k_x}+A_x)^2+(-i\partial_{k_y}+A_y)^2-\Omega(\bm{k})\right],\label{mom-space-landau}
\end{align}
where the Berry curvature is calculated by replacing $\bm{r}_i$ in Eq. (\ref{posi-berry}) with $\tilde{\bm{r}}_i$.
This is a Landau problem with an additional potential term $\Omega(\bm{k})$ on a momentum-space torus. A similar momentum-space Hamiltonian operator was discussed in terms of the momentum-space Landau level \cite{Claassen-Lee-Thomale-Qi-Devereaux-15,Lee-Claassen-Thomale-17}, though it does not have an exact zero mode.

\subsection{Numerical calculation in finite periodic system\label{periodicnotation}}
In most applications of band theory, a finite system with periodic boundary conditions is considered an approximation of an infinite system without boundaries.
The vortex function, however, contains the position operators, which are incompatible with the periodic boundary condition.
Nevertheless, the following method works well.

Let us consider the operators in the Chern-band eigenbasis $\{\ket{\bm{k}}\}$, where the band index is omitted.
Then, the lattice vortex function becomes an $N_{\rm unit}\times N_{\rm unit}$ matrix:
\begin{align}
    &PZP,PZ^*P\leftrightarrow\hat{Z},\hat{Z}^{\dagger},\notag\\
    &PZPZ^*P\leftrightarrow \hat{Z}\hat{Z}^{\dagger},\notag\\
    &[\hat{Z}]_{\bm{k},\bm{k}'}:=\bra{\bm{k}}Z\ket{\bm{k}'}.
\end{align}
Here we consider a system where the origin of the position coordinates is at the center, and periodic boundary conditions are imposed at both ends of the system.
Instead of operators defined on the infinite system, one can use $\hat{Z}$ and $\hat{Z}^{\dagger}$.
Roughly speaking, the eigenstate of $PZPZ^*P$ with the $m$-th smallest eigenvalue is localized on a (approximate) circle centered at the origin, with a radius proportional to $\sqrt{m}$.
Thus, when we focus on the eigenstates with small eigenvalues, the effect of the boundary does not matter.
One point to note is that there can exist small-eigenvalue states at the boundary of the system, where the position operators are discontinuous.
In fact, zero modes can appear at the corner of the system. Such artificial zero modes need to be removed by some method. In this paper, we introduce a small potential term at the corner to $\hat{Z}\hat{Z}^{\dagger}$.

\begin{figure}[]
\begin{center}
 \includegraphics[width=8cm,angle=0,clip]{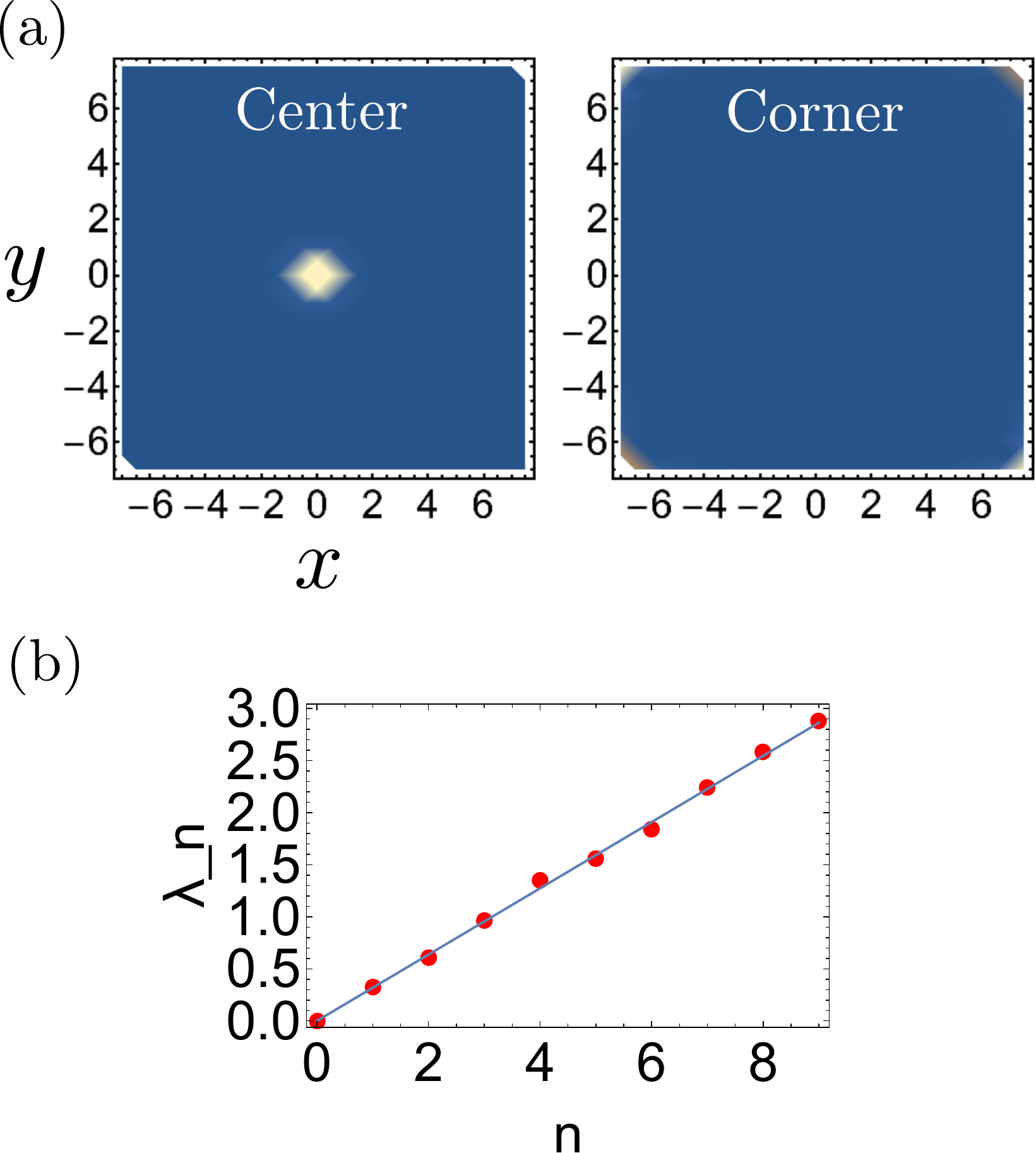}
 \caption{(a) Zero-modes of $PZPZ^*P$ in checkerboard lattice model. (b) The ten smallest eigenvalues of $\hat{Z}\hat{Z}^{\dagger}$, except for one quasi-zero mode. The blue line corresponds to $\lambda_n=n/\pi$.} 
 \label{fig1}
\end{center}
\end{figure}

As an example, we calculate the real-space configurations of the zero-modes of $PZPZ^*P$ for a model on the checkerboard lattice \cite{Neupert-Santos-Chamon-Mudry-11}.
This model is defined on a square lattice and includes two orbital degrees of freedom within the unit cell.
See Appendix for details.
We set $(\alpha_1,\alpha_2)=(1,i)$, $\tilde{\bm{r}}_1=(0.5,0)$, and $\tilde{\bm{r}}_2=(0,0.5)$, where the lattice constant is unity.
This virtual sublattice configuration was determined such that the lattice ``vortexability" is minimized \cite{okuma2024constructing}.
The system size $N_{\rm unit}$ is $15\times15$.
We find one (almost) exact zero mode ($\sim10^{-12}$) at the center and one quasi-zero mode ($\sim10^{-3}$) at the corner [Fig.\ref{fig1}(a)]. The latter zero mode is not essential for our problem. 

For the same setup, we also calculate the eigenvalues of $\hat{Z}\hat{Z}^{\dagger}$, except for one quasi-zero mode.
In the lowest Landau level, the eigenvalues of the counterpart of $PZPZ^*P$ are quantized as $\lambda_n=n/\pi$ because the system is described by the quantum harmonic oscillator. In terms of the eigenvalues, this model shows a similar behavior [Fig.\ref{fig1}(b)]. 
The eigenvalues of $PZPZ^*P$ and $PZZ^*P$ were also investigated in our previous work \cite{okuma2024constructing} and Ref. \cite{Claassen-Lee-Thomale-Qi-Devereaux-15,Lee-Claassen-Thomale-17}, respectively.
In our previous work \cite{okuma2024constructing}, we calculated the eigenvalues of Eq.(\ref{mom-space-landau}) under a discretization, and the values are slightly different from the eigenvalues of $\hat{Z}\hat{Z}^{\dagger}$.
The difference is expected to vanish in the infinite-volume limit.

\subsection{Coherent-like state on von Neumann lattice}
We will consider again the infinite system.
Actually, one can construct an infinite number of eigenstates of $PZ^*P$ because the following holds trivially:
\begin{align}
    \hat{z}^{\dagger}_{\bm{k}}[e^{-i\bm{k}\cdot\bm{R}}a_0(\bm{k})]&=[\hat{z}^{\dagger}_{\bm{k}}e^{-i\bm{k}\cdot\bm{R}}]a_0(\bm{k})\notag\\
    &=Z^*_{\bm{R}}[e^{-i\bm{k}\cdot\bm{R}}a_0(\bm{k})],\\
    PZ^*P\ket{\zeta_{\bm{R}}}&=Z^*_{\bm{R}}\ket{\zeta_{\bm{R}}},\\
    \ket{\zeta_{\bm{R}}}&\propto\int\frac{d^2k}{(2\pi)^2}e^{-i\bm{k}\cdot\bm{R}}a_0(\bm{k})\ket{\bm{k}},
\end{align}
where $Z_{\bm{R}}=\alpha_1X+\alpha_2Y$.
This is an analogy of the coherent states, which are the eigenstates of the annihilation operator. Strictly speaking, $[\hat{z}^{\dagger}_{\bm{k}},\hat{z}_{\bm{k}}]=2\Omega(\bm{k})$, and $(\hat{z}^{\dagger}_{\bm{k}},\hat{z}_{\bm{k}})$ can act as annihilation and creation operators only when the Berry curvature is a constant of momentum.
We will call these states the coherent-like states.
In the case of the true coherent states, complete subsets of the (overcomplete) set of the coherent states exist.
These states are characterized by a two-dimensional translation-invariant lattice, called von Neumann lattice \cite{von2018mathematical,perelomov2002completeness, bargmann1971completeness}.
Note that these subsets are complete but not orthogonal.
The notion of the von Neumann lattice was discussed in quantum Hall physics \cite{imai1990field,ishikawa1992field,ishikawa1999field}, where the lowest Landau level is described by creation and annihilation operators.
As an analogy, we regard the set $\{\ket{\zeta_{\bm{R}}}\}$ as a natural generalization of the coherent states on the von Neumann lattice.
In our case, the von Neumann lattice is nothing but the periodic lattice on which the model is defined.

The coherent-like state is characterized by a coefficient $a_0(\bm{k})$.
Interestingly, $a_0(\bm{k})$ has a zero point $\bm{k}_0$ in the Brillouin zone.
If we add a constant shift term to the virtual sublattice positions $\{\tilde{\bm{r}}_i\}$,
this zero point moves in the Brillouin zone.
This is because the constant shift means the flux insertion to the noncontractible loops of the momentum-space torus.
In the case of quantum Hall physics, the corresponding coefficient has a zero point at $\bm{k}=\bm{0}$. To mimic this feature, if $\bm{k}_0\neq\bm{0}$ in our problem, we add a constant complex number to the lattice vortex function such that the zero point of $a_0(\bm{k})$ exists at $\bm{k}=\bm{0}$.
Note that the addition of the constant changes the eigenspectrum and eigenstates of $PZPZ^*P$ because there is no translation invariance inside the unit cell.

\begin{figure}[]
\begin{center}
 \includegraphics[width=8cm,angle=0,clip]{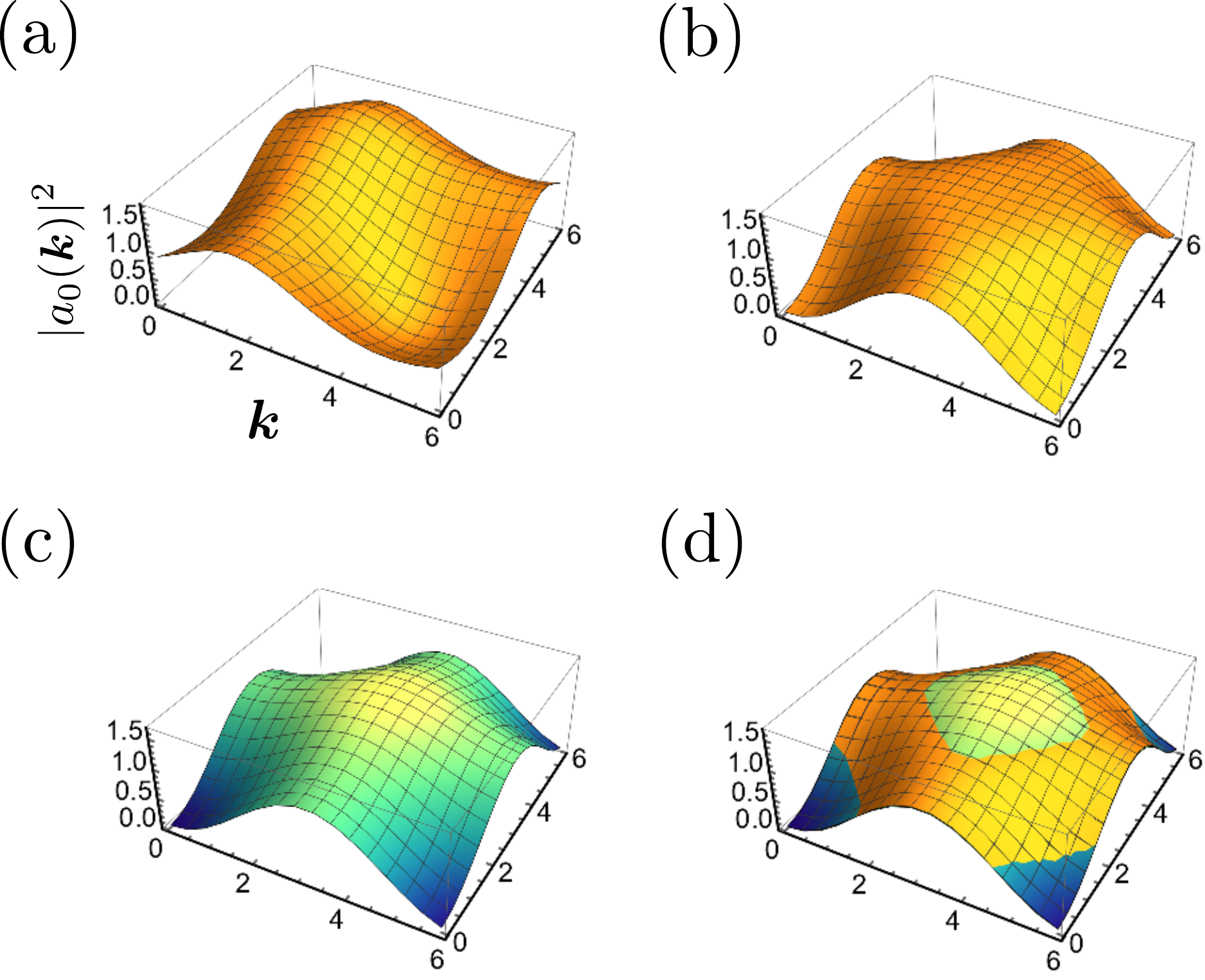}
 \caption{Distributions $|a_0(\bm{k})|^2$ for (a) $\tilde{\bm{r}}_1=(0.25,-0.25),~\tilde{\bm{r}}_2=(-0.25,0.25)$, (b) $\tilde{\bm{r}}_1=(0.5,0),~\tilde{\bm{r}}_2=(0,0.5)$. The counterpart in the lowest Landau level is shown in (c). The comparison between (b) and (c) is shown in (d).}
 \label{fig2}
\end{center}
\end{figure}

As an example, we again consider the checkerboard-lattice model.
We calculate the distribution $|a_0(\bm{k})|^2$ for (a) $\tilde{\bm{r}}_1=(0.25,-0.25),~\tilde{\bm{r}}_2=(-0.25,0.25)$, (b) $\tilde{\bm{r}}_1=(0.5,0),~\tilde{\bm{r}}_2=(0,0.5)$.
Here, the constant shift is $(0.25,0.25)$.
In the case (a), the zero point $\bm{k}_0$ is not at the origin, while it is at the origin in the case (b) [Fig.\ref{fig2}(a,b)].
In the case of the lowest Landau level, $\alpha(\bm{p})$ in Ref.\cite{ishikawa1999field} corresponds to $|a_0(\bm{k})|^2$:
\begin{align}
    \alpha(\bm{p})&=[\beta(\bm{p})]^*\beta(\bm{p}),\\
    \beta(\bm{p})&=(2\mathrm{Im}\tau)^{\frac{1}{4}}e^{i\frac{\tau}{4\pi}p_y^2}\vartheta_1(\frac{p_x+\tau p_y}{2\pi}|\tau),
\end{align}
where $\bm{p}$ is the momentum in the ``Brillouin zone" of the von Neumann lattice, $\tau=i$ for the square lattice, and $\vartheta_1$ is the elliptic theta function.
We compare $|a_0(\bm{k})|^2$ in the case (b) with $\alpha(\bm{p})$ [Fig.\ref{fig2}(c, d)].
In terms of the distribution $|a_0(\bm{k})|^2$, the checkerboard-lattice model is very similar to the lowest Landau level.

\begin{figure}[]
\begin{center}
 \includegraphics[width=6cm,angle=0,clip]{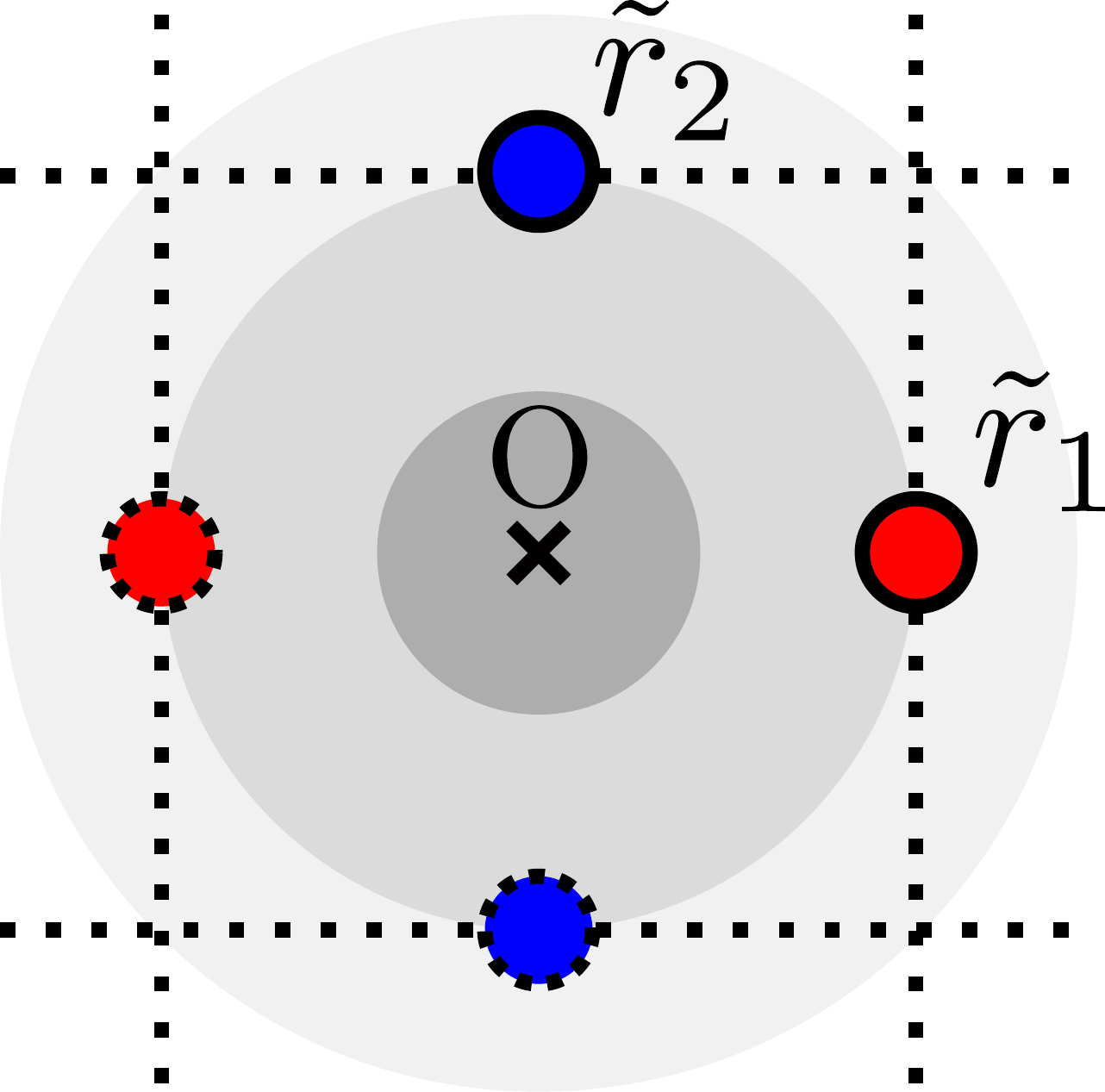}
 \caption{Von Neumann lattice center for checkerboard-lattice model.}
 \label{fig3}
\end{center}
\end{figure}

It is interesting to note that we can define the ``center" of the coherent-like state by setting $\tilde{\bm{r}}$'s such that $a_0(\bm{k}=\bm{0})=0$, while the Wannier center or the polarization is not well defined in Chern insulators \cite{coh2009electric}.
We call it the von Neumann lattice center.
If we regard $\tilde{\bm{r}}$'s as real sublattice positions, the origin $(0,0)$ for the sublattice positions is uniquely determined for a given Chern insulator, up to the lattice constant. For example, the relationship between the von Neumann lattice center (O) and $\tilde{\bm{r}}$'s for the checkerboard-lattice model is shown in Fig. \ref{fig3}.

\subsection{Biorthogonal basis}
The coherent-like states are not orthogonal to each other, as in the case of the coherent states.
Instead of an orthogonal basis, we construct the biorthogonal basis for the coherent-like states.
We again consider the finite and periodic system in Sec. \ref{periodicnotation}.
We assume that $a_0(\bm{k})$ is obtained as the zero mode of $\hat{Z}\hat{Z}^{\dagger}$, and that $\bm{k}_0$ is adjusted to be $\bm{0}$, i.e., $a_0(\bm{k}=0)=0$, by adding a complex constant to the lattice vortex function.
Although there are some differences, the following construction extends the formalism for the quantum Hall effect in Ref.\cite{imai1990field} to Chern insulators.

In the finite system, a coherent-like state is defined as
\begin{align}
    \ket{\zeta_{\bm{R}}}=\sum_{\bm{k}}\frac{e^{-i\bm{k}\cdot\bm{R}}}{\sqrt{N_{\rm unit}}}a_{0}(\bm{k})\ket{\bm{k}}.
\end{align}
For convenience, we impose the following normalization condition on $\ket{\zeta_{\bm{R}}}$:
\begin{align}
    \bra{\zeta_{\bm{R}}}\zeta_{\bm{R}}\rangle=\frac{1}{N_{\rm unit}}\sum_{\bm{k}}|a_{0}(\bm{k})|^2=1.
\end{align}
The coherent-like state satisfies the following eigenvalue equation:
\begin{align}
    PZ^*P\ket{\zeta_{\bm{R}}}&=[Z_{\bm{R}}]^*\ket{\zeta_{\bm{R}}},\notag\\
    \Leftrightarrow\bra{\zeta_{\bm{R}}}PZP&=Z_{\bm{R}}\bra{\zeta_{\bm{R}}}.
\end{align}
Thus, $\bra{\zeta_{\bm{R}}}$ is a left eigenstate of $PZP$.
We define the conjugate basis $\{\ket{Z_{\bm{R}}}\}$:
\begin{align}
    \ket{Z_{\bm{R}}}=\sum_{\bm{k}\neq\bm{0}}\frac{e^{-i\bm{k}\cdot\bm{R}}}{\sqrt{N_{\rm unit}}}\frac{1}{[a_{0}(\bm{k})]^*}\ket{\bm{k}}.
\end{align}
Since $a_0$ is $\bm{0}$ at $\bm{k=0}$, $\bm{k}=0$ is omitted in the summation.
The sets $\{\ket{\zeta_{\bm{R}}}\}$ and $\{\ket{Z_{\bm{R}}}\}$ satisfy the following nearly biorthogonal condition:
\begin{align}
    \bra{\zeta_{\bm{R}}}Z_{\bm{R}'}\rangle=\frac{1}{N_{\rm unit}}\sum_{\bm{k}\neq\bm{0}} e^{i\bm{k}\cdot (\bm{R}-\bm{R}')}=
    \delta_{\bm{R}\bm{R}'}-\frac{1}{N_{\rm unit}}.
\end{align}
In the infinite-volume limit, we have the biorthogonal basis.

The system size should be finite in a numerical calculation, and the above basis is not useful.
Instead, we use the following basis functions:
\begin{align}
    &\ket{\zeta'_{\bm{R}}}=\ket{\zeta_{\bm{R}}}+\frac{A}{\sqrt{N_{\rm unit}}}\ket{\bm{k}=0},\\
    &\ket{Z'_{\bm{R}}}=\ket{Z_{\bm{R}}}+\frac{1}{A\sqrt{N_{\rm unit}}}\ket{\bm{k}=0},
\end{align}
where $A$ is an $\mathcal{O}(1)$ constant.
These states satisfy
\begin{align}
    \bra{\zeta'_{\bm{R}}}Z'_{\bm{R}'}\rangle=
    \delta_{\bm{R}\bm{R}'}-\frac{1}{N_{\rm unit}}+\frac{1}{N_{\rm unit}}=\delta_{\bm{R}\bm{R}'}.\label{biorthogonalcondition}
\end{align}
In Ref.\cite{imai1990field}, a similar treatment is performed for the Landau-level counterpart.
In the following, we set $A=1$ for simplicity. The results of the numerical calculations in the latter part are almost independent of $A$.

In summary, the biorthogonal basis consists of the following functions:
\begin{align}
\ket{\zeta'_{\bm{R}}}&=\sum_{\bm{k}}\frac{e^{-i\bm{k}\cdot\bm{R}}}{\sqrt{N_{\rm unit}}}a'_{0}(\bm{k})\ket{\bm{k},\alpha},\\
\bra{\zeta'_{\bm{R}}}\bm{r},i\rangle&=\sum_{\bm{k}}\frac{e^{i\bm{k}\cdot\bm{R}}}{\sqrt{N_{\rm unit}}}\frac{e^{-i\bm{k}\cdot\bm{r}}}{\sqrt{N_{\rm unit}}}[a'_0(\bm{k})]^*[u_{\bm{k}}(i)]^*\notag\\
&=:\sum_{\bm{k}}\frac{e^{i\bm{k}\cdot\bm{R}}}{\sqrt{N_{\rm unit}}}\frac{e^{-i\bm{k}\cdot\bm{r}}}{\sqrt{N_{\rm unit}}}\bar{v}_{\bm{k}}(i),\\
\ket{Z'_{\bm{R}}}&=\sum_{\bm{k}}\frac{e^{-i\bm{k}\cdot\bm{R}}}{\sqrt{N_{\rm unit}}}\frac{1}{[a'_{0}(\bm{k})]^*}\ket{\bm{k},\alpha}.\\
    \bra{\bm{r},i} Z'_{\bm{R}}\rangle&=\sum_{\bm{k}}\frac{e^{-i\bm{k}\cdot\bm{R}}}{\sqrt{N_{\rm unit}}}\frac{e^{i\bm{k}\cdot\bm{r}}}{\sqrt{N_{\rm unit}}}\frac{1}{[a'_0(\bm{k})]^*}u_{\bm{k}}(i)\notag\\
    &=:\sum_{\bm{k}}\frac{e^{-i\bm{k}\cdot\bm{R}}}{\sqrt{N_{\rm unit}}}\frac{e^{i\bm{k}\cdot\bm{r}}}{\sqrt{N_{\rm unit}}}v_{\bm{k}}(i).
\end{align}
where 
\begin{align}
    a'_0(\bm{k})=
    \begin{cases}
        a_0(\bm{k})&\bm{k}\neq\bm{0}\\
        1&\bm{k}=\bm{0}
    \end{cases}~~~~.
\end{align}

\subsection{Creation and annihilation operators on von Neumann lattice}
By using the biorthogonal basis that consists of $\{\ket{\zeta'_{\bm{R}}}\}$ and $\{\ket{Z'_{\bm{R}}}\}$, we define the creation and annihilation operators on the von Neumann lattice:
\begin{align}
\bar{C}_{\bm{R}}=\sum_{\bm{r},i}\bra{\bm{r},i} Z'_{\bm{R}}\rangle c^{\dagger}_{\bm{r},i},~C_{\bm{R}}=\sum_{\bm{r},i}\bra{\zeta'_{\bm{R}}}\bm{r},i\rangle c_{\bm{r},i}.
\end{align}
Although $\bar{C}_{\bm{R}}$ is NOT the Hermitian conjugate of $C_{\bm{R}}$, they satisfy the fermionic anticommutation relations:
\begin{align}
    \{\bar{C}_{\bm{R}}, C_{\bm{R}'}\}=\delta_{\bm{R}\bm{R}'}, \{\bar{C}_{\bm{R}}, \bar{C}_{\bm{R}'}\}=\{C_{\bm{R}}, C_{\bm{R}'}\}=0.
\end{align}
The first relation is derived by using the biorthogonal condition (\ref{biorthogonalcondition}).
Under the projection onto the Chern band, the original fermionic operators are approximated as
\begin{align}
c^{\dagger}_{\bm{r},i}\simeq\sum_{\bm{R}}\bra{\zeta'_{\bm{R}}}\bm{r},i\rangle \bar{C}_{\bm{R}},~
    c_{\bm{r},i}\simeq\sum_{\bm{R}}\bra{\bm{r},i} Z'_{\bm{R}}\rangle C_{\bm{R}}.\label{coherentexpansion}
\end{align}
In the following section, we consider an application of the biorthogonal basis to the FCI search.

\section{Application of biorthogonal basis to fractional Chern insulator\label{applicationsection}}
In this section, we introduce a composite fermionization based on the two-dimensional Jordan-Wigner transformation for creation and annihilation operators on the von Neumann lattice.
We formulate a composite-fermion mean-field theory of $C=1$ Chern-insulator models with the filling factor $\nu=1/q$, under a short-range repulsive interaction.
As examples, we numerically calculate the composite-fermion band structures of FCI candidates with a filling factor $\nu=1/3$.
Throughout this section, we assume that $(\alpha_1,\alpha_2)=(1,i)$, which means that the unit cell is square, and the von Neumann lattice is identical to the square lattice.
We set the lattice constant to unity.

\subsection{2D Jordan-Wigner transformation for biorthogonal basis}
Reference \cite{fradkin1989jordan} introduced a two-dimensional Jordan-Wigner transformation on a lattice model. This transformation describes flux attachment in the second quantization formalism.
As a simple extension to biorthogonal basis, we consider the two-dimensional Jordan-Wigner transformation on von Neumann lattice:
\begin{align}
    &\bar{\beta}_{\bm{R}}=\prod_{\bm{r}\neq\bm{R}}(Z_{\bm{R}}-Z_{\bm{r}})^{fn_{\bm{r}}}\bar{C}_{\bm{R}},\notag\\
    &\beta_{\bm{R}}=\prod_{\bm{r}\neq\bm{R}}(Z_{\bm{R}}-Z_{\bm{r}})^{-fn_{\bm{r}}}C_{\bm{R}},
\end{align}
where $f$ is an even integer representing the number of attached flux quanta (in our case, $f=q-1=1/\nu-1$), and $n_{\bm{r}}=\bar{C}_{\bm{r}}C_{\bm{r}}$ is the number operator on $\bm{r}$.
The new creation and annihilation operators, $(\bar{\beta},\beta)$, satisfy the fermionic commutation relation.
Instead of the phase factor $e^{i\theta(\bm{R},\bm{r})}$ in the original version, we use $Z_{\bm{R}}-Z_{\bm{r}}$.
Basically, similar transformations can be considered on any basis.
The reason we chose the biorthogonal basis on the von Neumann lattice is that it is diagonal for the projected vortex function in the infinite-volume limit.

The Jordan Wigner transformation is useful for rewriting the quadratic terms of creation and annihilation operators.
 For $\bm{R}=\bm{R}'$, $\bar{\beta}_{\bm{R}}\beta_{\bm{R}'}=\bar{C}_{\bm{R}}C_{\bm{R}'}$, and for $\bm{R}\neq\bm{R}'$,
\begin{align}
    \bar{\beta}_{\bm{R}}\beta_{\bm{R}'}&=\bar{C}_{\bm{R}}\prod_{\bm{r}\neq\bm{R}}(Z_{\bm{R}}-Z_{\bm{r}})^{fn_{\bm{r}}}\notag\\&\prod_{\bm{r}\neq\bm{R}'}(Z_{\bm{R}'}-Z_{\bm{r}})^{-fn_{\bm{r}}}
    C_{\bm{R}'}\notag\\
    &=\bar{C}_{\bm{R}}\prod_{\bm{r}\neq\bm{R},\bm{R}'}\left(\frac{Z_{\bm{R}}-Z_{\bm{r}}}{Z_{\bm{R}'}-Z_{\bm{r}}}\right)^{fn_r}\notag\\&(Z_{\bm{R}}-Z_{\bm{R}'})^{fn_{\bm{R}'}}(Z_{\bm{R}'}-Z_{\bm{R}})^{-fn_{\bm{R}}}C_{\bm{R}'}\notag\\
    &=\bar{C}_{\bm{R}}\prod_{\bm{r}\neq\bm{R},\bm{R}'}\left(\frac{Z_{\bm{R}}-Z_{\bm{r}}}{Z_{\bm{R}'}-Z_{\bm{r}}}\right)^{fn_r}\notag\\&(Z_{\bm{R}}-Z_{\bm{R}'})^{f(n_{\bm{R}'}-n_{\bm{R}})}C_{\bm{R}'}
\end{align}
Since $f$ is even and $n_{\bm{R}}$ is 0 or 1 in the Fock space, the factor $(-1)^{-fn_{\bm{R}}}$ is unity. 

\subsection{Mean-field approximation for number operator}
In the absence of the Chern-Simons gauge field, the above transformation does not simplify the problem of strongly correlated systems.
We here introduce a translation-invariant mean-field approximation for the number operator:
\begin{align}
    n_{\bm{R}}=\nu.\label{uniformapprox}
\end{align}
This bold approximation is based on the assumption that the ground state is gapped and fluctuations are suppressed.
The approximation of replacing the particle density with the c-number is made in a previous study \cite{maiti2019fermionization} using the 2D Jordan-Wigner transformation.
We also assume that the topological order does not break translation symmetry, and the expectation value of the particle density is uniform.
Under the approximation, we obtain
\begin{align}
    \bar{\beta}_{\bm{R}}\beta_{\bm{R}'}=&\left[\bar{C}_{\bm{R}}A(\bm{R})\right]\exp\left(\sum_{\bm{r}\neq\bm{R},\bm{R}'} if\nu \Delta\theta(\bm{R},\bm{R}':\bm{r})\right)\notag\\
    &\left[A^{-1}(\bm{R}')C_{\bm{R}'}\right],
\end{align}
where 
\begin{align}
    A(\bm{R})&:=\prod_{\bm{r}\neq\bm{R}}|Z_{\bm{R}}-Z_{\bm{r}}|^{f\nu}\\
    \Delta\theta(\bm{R},\bm{R}':\bm{r})&:=\arg \left(\frac{Z_{\bm{R}}-Z_{\bm{r}}}{Z_{\bm{R}'}-Z_{\bm{r}}}\right).
\end{align}
Because a similarity transformation does not change the anticommutation relation, we use the following new operators:
\begin{align}
    \gamma_{\bm{R}}=A(\bm{R})\beta_{\bm{R}},~\bar{\gamma}_{\bm{R}}=A^{-1}(\bm{R})\bar{\beta}_{\bm{R}}.
\end{align}
By evaluating the summation, we obtain
\begin{align}
    \bar{\gamma}_{\bm{R}}\gamma_{\bm{R}'}&=\bar{C}_{\bm{R}}e^{i\phi(\bm{R},\bm{R}')}C_{\bm{R}'},\\
    \phi(\bm{R},\bm{R}')&=f\nu \pi \hat{e}_z\cdot (\bm{R}\times\bm{R}')\notag\\
    &=f\nu (R_xR'_y-R_yR'_x).\label{symmetric}
\end{align}
This phase factor represents an emergent homogeneous magnetic field induced by interaction. Finally, the constraint for this approximation is rewritten as
\begin{align}
    \bar{\gamma}_{\bm{R}}\gamma_{\bm{R}}=\nu.\label{constraint}
\end{align}

\subsection{Landau gauge, magnetic unit cell, and magnetic Brillouin zone}
Instead of the symmetric gauge (\ref{symmetric}), we adopt the Landau gauge:
\begin{align}
    \phi(\bm{R},\bm{R}')=\pi f\nu (R_x+R_{x'})(R_y-R_{y'}).
\end{align}
For $\nu=1/q$, one magnetic unit cell for composite fermions corresponds to $q$ original unit cells.
We here introduce a magnetic lattice vector $\bm{d}\in\mathbb{Z}^2$ via
\begin{align}
    \bm{R}=(qd_x+n,d_y),
\end{align}
where $n=0,1,\cdots,q-1$ describes original unit-cell vectors in the magnetic unit cell.
Using the magnetic lattice vector, the phase is rewritten as
\begin{align}
    \phi(\bm{R},\bm{R}')&=\pi f\left[(d_x+d_{x'})+\frac{(n+n')}{q}\right](d_y-d_{y'})\notag\\
    &\equiv\frac{\pi f}{q}(n+n')(d_y-d_{y'})\pmod {2\pi}\notag\\
    &=: p^{nn'}_y (d_y-d_{y'}).
\end{align}
In the second line, we have used the facts that $f$ is even and $q\nu=1$.
In summary, we obtain
\begin{align}
    \bar{C}_{\bm{d},n}C_{\bm{d}',n'}=\bar{\gamma}_{\bm{d},n}e^{-ip^{nn'}_y(d_y-d_y')}\gamma_{\bm{d}',n'},\label{correspondence}
\end{align}
where the subscript has been replaced from $\bm{R}$ to $(\bm{d},n)$.

The new Brillouin zone, which corresponds to the enlarged unit cell, undergoes folding. This is nothing but the magnetic Brillouin zone for composite fermions.
We change the scaling in the $x$-direction to keep the size of the Brillouin zone:
\begin{align}
    &\bm{K}=([qk_x],k_y),\\
    &\bm{K}\in [0,2\pi)\times[0,2\pi),\\
    &qk_x=K_x+2\pi m_k\in [0,2\pi q),
\end{align}
where $[qk_x]\equiv qk_x~(\mathrm{mod}~2\pi)$, and $m_{k}=0,1,\cdots,q-1$. We will denote the momentum within the magnetic Brillouin zone using capital letters.
The Fourier transforms of old and new fermionic operators are defined as
\begin{align}
    &\bar{C}_{\bm{K},n}=\frac{1}{\sqrt{N_{\rm unit}/q}}\sum_{\bm{d}}e^{i\bm{K}\cdot\bm{d}}\bar{C}_{\bm{d},n},\label{ckn}\\
    &C_{\bm{K},n}=\frac{1}{\sqrt{N_{\rm unit}/q}}\sum_{\bm{d}}e^{-i\bm{K}\cdot\bm{d}}C_{\bm{d},n},\label{cknn}\\
    &\bar{\gamma}_{\bm{K},n}=\frac{1}{\sqrt{N_{\rm unit}/q}}\sum_{\bm{d}}e^{i\bm{K}\cdot\bm{d}}\bar{\gamma}_{\bm{d},n},\\
    &\gamma_{\bm{K},n}=\frac{1}{\sqrt{N_{\rm unit}/q}}\sum_{\bm{d}}e^{-i\bm{K}\cdot\bm{d}}\gamma_{\bm{d},n}.
\end{align}
The correspondence (\ref{correspondence}) is rewritten in momentum space as
\begin{align}
    &\bar{C}_{\bm{K},n}C_{\bm{K}',n'}=\left(\frac{1}{\sqrt{N_{\rm unit}/q}}\right)^2\sum_{\bm{d},\bm{d}'}e^{i\bm{K}\cdot\bm{d}}e^{-i\bm{K}'\cdot\bm{d}'}\bar{C}_{\bm{d}n}C_{\bm{d}'n'}\notag\\
    &\rightarrow\notag\\
    &\left(\frac{1}{\sqrt{N_{\rm unit}/q}}\right)^2\sum_{\bm{d},\bm{d}'}e^{i\bm{K}\cdot\bm{d}}e^{-i\bm{K}'\cdot\bm{d}'}e^{-i\bm{P}^{nn'}\cdot(\bm{d}-\bm{d}')}\bar{\gamma}_{\bm{d}n}\gamma_{\bm{d}'n'}\notag\\
    &=\bar{\gamma}_{\bm{K}-\bm{P}^{nn'},n}\gamma_{\bm{K}'-\bm{P}^{nn'},n'},
\end{align}
where $\bm{P}^{nn'}=\bm{p}^{nn'}=(0,p_y^{nn'})$.

\subsection{Density operator in composite fermions}
Our goal is to rewrite the many-body Hamiltonian in terms of the composite fermions.
Since the interaction Hamiltonian consists of density operators, we consider the projection of the density operator (\ref{densityoperator}) onto the Chern band and expand it in the biorthogonal basis:
\begin{align}
    &\tilde{\rho}_{\bm{q},i}:=\mathcal{P}\rho_{\bm{q},i}\mathcal{P}=\mathcal{P}\sum_{\bm{r}}e^{-i\bm{q}\cdot \bm{r}}c^{\dagger}_{\bm{r},i}c_{\bm{r},i}\mathcal{P}\notag\\
    =&\sum_{\bm{r},\bm{R},\bm{R}'}e^{-i\bm{q}\cdot \bm{r}}\bra{\zeta'_{\bm{R}}}\bm{r},i\rangle\bra{\bm{r},i} Z'_{\bm{R}'}\rangle\bar{C}_{\bm{R}}C_{\bm{R}'},
\end{align}
where $\mathcal{P}$ is the projection operator acting on the creation and annihilation operators. By using the momentum-space expression of the biorthogonal basis, we obtain
\begin{align}
    &\tilde{\rho}_{\bm{q},i}=\frac{1}{N_{\rm unit}^2}\sum_{\bm{R},\bm{R}'}\bar{C}_{\bm{R}}C_{\bm{R}'}\sum_{\bm{k},\bm{k}'}\sum_{\bm{r}}e^{i(\bm{k}'-\bm{k}-\bm{q})\cdot\bm{r}}\notag\\
    &e^{i\bm{k}\cdot\bm{R}}e^{-i\bm{k}'\cdot\bm{R}'}
    \bar{v}_{\bm{k}}(i)v_{\bm{k}'}(i)\notag\\
    =&\frac{1}{N_{\rm unit}}\sum_{\bm{R},\bm{R}'}\bar{C}_{\bm{R}}C_{\bm{R}'}\sum_{\bm{k}}e^{i\bm{k}\cdot\bm{R}}e^{-i(\bm{k}+\bm{q})\cdot\bm{R}'}
    \bar{v}_{\bm{k}}(i)v_{\bm{k}+\bm{q}}(i).
\end{align}
By using Eqs. (\ref{ckn}) and (\ref{cknn}), we can perform Fourier transforms corresponding to the magnetic Brillouin zone:
\begin{align}
    &\tilde{\rho}_{\bm{Q},m_q,i}:=\tilde{\rho}_{\bm{q},i}\notag\\
    &=\frac{1}{q}\sum_{\substack{n,n',\\\bm{K},m_k}}e^{ink_x}e^{-in'(k_x+q_x)}\bar{v}_{\bm{k}}(i)v_{\bm{k}+\bm{q}}(i)\bar{C}_{\bm{K},n}C_{\bm{K}+\bm{Q},n'},
\end{align}
where $\bm{k}=((K_x+2\pi m_k)/q,K_y)$, and $\bm{q}=((Q_x+2\pi m_q)/q,Q_y)$ are the momenta in the original Brillouin zone. 

Finally, we perform the composite fermionization with mean-field approximation:
\begin{align}
    &\tilde{\rho}_{\bm{Q},m_q,i}
    \notag\\
    \rightarrow&\frac{1}{q}\sum_{\substack{n,n',\\\bm{K},m_k}}e^{ink_x}e^{-in'(k_x+q_x)}\bar{v}_{\bm{k}}(i)v_{\bm{k}+\bm{q}}(i)\notag\\
    &\bar{\gamma}_{\bm{K}-\bm{P}^{nn'},n}\gamma_{\bm{K}+\bm{Q}-\bm{P}^{nn'},n'}\notag\\
    =&\frac{1}{q}\sum_{\substack{n,n',\\\bm{K},m_k}}e^{ink_x}e^{-in'(k_x+q_x)}\bar{v}_{\bm{k}+\bm{p}^{nn'}}(i)v_{\bm{k}+\bm{p}^{nn'}+\bm{q}}(i)\notag\\
    &\bar{\gamma}_{\bm{K},n}\gamma_{\bm{K}+\bm{Q},n'}.
\end{align}

\subsection{Interaction Hamiltonian of composite fermions on biorthogonal basis}
In the following, we focus on the correlation and neglect the band dispersion of the kinetic part. The topological and geometrical effects of the band structure are included via the projection of the interaction Hamiltonian onto the Chern band.
We consider the following Hamiltonian:

\begin{widetext}
\begin{align}
    H_{\rm eff}=&\frac{1}{2N_{\rm unit}}\sum_{i,j}\sum_{\bm{q}}V^{ij}(-\bm{q})\tilde{\rho}_{\bm{q},i}\tilde{\rho}_{-\bm{q},j}
    =\frac{1}{2N_{\rm unit}}\sum_{i,j}\sum_{\bm{Q},m_q}V^{ij}(-\bm{Q},m_q)\tilde{\rho}_{\bm{Q},m_q,i}\tilde{\rho}_{-\bm{Q},-m_q,j}\notag\\
\rightarrow&\sum_{n,n',n'',n'''}\sum_{\bm{K},\bm{K}',\bm{Q}}
    V^{n,n',n'',n'''}_{\bm{K},\bm{K}+\bm{Q},\bm{K}',\bm{K}'-\bm{Q}}~\bar{\gamma}_{\bm{K},n}\gamma_{\bm{K}+\bm{Q},n'}\bar{\gamma}_{\bm{K}',n''}\gamma_{\bm{K}'-\bm{Q},n'''},\label{mapping}\\
    V^{n,n',n'',n'''}_{\bm{K},\bm{K}+\bm{Q},\bm{K}',\bm{K}'-\bm{Q}}=&\frac{1}{2N_{\rm unit}}\sum_{i,j}\sum_{m_{q}}V^{ij}(-\bm{Q},-m_q)\rho(i,n,n',\bm{K},\bm{Q},m_q)\rho(j,n'',n''',\bm{K}',-\bm{Q},-m_q),\\
    \rho(i,n,n',\bm{K},\bm{Q},m_q)=&\frac{1}{q}\sum_{m_k}e^{i\frac{n}{q}(K_x+2\pi m_k)}e^{-i\frac{n'}{q}(K_x+Q_x+2\pi(m_k+m_q))}\notag\\
    &\bar{v}_{\left(\frac{K_x+2\pi m_k}{q},K_y+P^{nn'}_y\right)}(i)v_{\left(\frac{K_x+Q_x+2\pi (m_k+m_q)}{q},K_y+Q_y+P^{nn'}_y\right)}(i),\\
    V^{ij}(-\bm{Q},-m_q)=&V^{ij}(-\bm{q})=\sum_{\Delta\bm{r}}e^{i\left(\frac{Q_x+2\pi m_q}{q},Q_y\right)\cdot(\Delta\bm{r})}V^{ij}(\Delta \bm{r}).
\end{align}
\end{widetext}

\subsection{Non-Hermitian Hartree-Fock mean-field theory}
In the following, we apply the Hartree-Fock mean-field theory to our problem.
In this framework, the effective Hamiltonian $H_{\rm eff}$ is treated approximately as a quadratic Hamiltonian:
\begin{align}
    H_{\rm eff}\simeq \sum_{\bm{K},a,b}&H^{ab}_{\bm{K}}~\bar{\gamma}_{\bm{K},a}\gamma_{\bm{K},b},\\
    H^{ab}_{\bm{K}}=\sum_{\bm{K}',c,d}&\left(V^{a,b,c,d}_{\bm{K},\bm{K},\bm{K}',\bm{K}'}+V^{c,d,a,b}_{\bm{K}',\bm{K}',\bm{K},\bm{K}}\right.\notag\\
    &\left.-V^{c,b,a,d}_{\bm{K}',\bm{K},\bm{K},\bm{K}'}-V^{a,d,c,b}_{\bm{K},\bm{K}',\bm{K}',\bm{K}}\right)\langle \bar{\gamma}_{\bm{K}',c}\gamma_{\bm{K}',d}\rangle\notag\\
    +\sum_{\bm{K}',c}& V^{a,c,c,b}_{\bm{K},\bm{K}',\bm{K}',\bm{K}}~~~.\label{hamexpression}
\end{align}
Here we assume that the magnetic Brillouin zone is irreducible, and the correlation function $\langle\bar{\gamma}_{\bm{K,a}}\gamma_{\bm{K',b}}\rangle$ is non-zero only for $\bm{K}=\bm{K}'$.
The elements $\langle\bar{\gamma}_{\bm{K,a}}\gamma_{\bm{K,b}}\rangle$ should be determined selfconsistently, as shown below.
The last constant term arises when $\bar{\gamma}_1\gamma_2\bar{\gamma}_3\gamma_4$ is rearranged to be $\bar{\gamma}_1\bar{\gamma}_3\gamma_4\gamma_2$ before performing the approximation.
Unlike the conventional problems, the matrix representation of our quadratic Hamiltonian becomes a $q\times q$ non-Hermitian matrix.
Due to the approximation, there is no direct reason for the energy spectrum to be real, while we expect that it approaches the real axis if the approximation is valid.

In the mean-field theory, the correlation functions $\langle \bar{\gamma}_{\bm{K},a}\gamma_{\bm{K},b}\rangle$ are obtained by self-consistently performing the following three steps.
\begin{enumerate}
    \item Diagonalize $\hat{H}_{\bm{K}}$ by a matrix $\hat{N}_{\bm{K}}$ (not always unitary):
    \begin{align}
        &\hat{N}^{-1}_{\bm{K}}\hat{H}_{\bm{K}}\hat{N}_{\bm{K}}=\hat{E}_{\bm{K}},\\
        &H_{\rm eff}\simeq \sum_{\bm{K},a,b}\bar{\gamma}_{\bm{K},a}~H^{ab}_{\bm{K}}~\gamma_{\bm{K},b}=\sum_{\bm{K},\alpha,\beta} E_{\bm{K},\alpha}\bar{\gamma}_{\bm{K},\alpha}\gamma_{\bm{K},\alpha}.
    \end{align}
    Here,
    \begin{align}
    \bar{\gamma}_{\bm{K},\alpha}=\sum_{a}\bar{\gamma}_{\bm{K},a}N^{a\alpha}_{\bm{K}}, ~\gamma_{\bm{K},\alpha}=\sum_{a}[N^{-1}_{\bm{K}}]^{\alpha a}\gamma_{\bm{K},a}.
    \end{align}
    \item Calculate the matrix $\hat{C}_{\bm{K}}$ whose matrix element is given by $\langle \bar{\gamma}_{\bm{K},a}\gamma_{\bm{K},b}\rangle$:
    \begin{align}
        \hat{C}_{\bm{K}}&=[\hat{N}_{\bm{K}}^{-1}]^{T}\hat{F}_{\bm{K}}\hat{N}_{\bm{K}}^{T},\\
        [\hat{F}_{\bm{K}}]^{\alpha\beta}&=
        \begin{cases}
            1&\alpha=\beta=\mathrm{lowest}~\mathrm{band},\\
            0&\mathrm{otherwize}.
        \end{cases}
    \end{align}
    \item Calculate new $\hat{H}_{\bm{K}}$ by using $\{\hat{C}_{\bm{K}'}\}$. Then, back to the first step.
\end{enumerate}
The definition of the ``lowest" band in step 2 is nontrivial because the energy spectrum is not always real.
We will order the energies according to the real part.
From Eq. (\ref{constraint}), the following constraint should be satisfied:
\begin{align}
    \nu&=\langle\bar{\gamma}_{\bm{d},a}\gamma_{\bm{d},a}\rangle\notag\\
    &=\left(\frac{1}{\sqrt{N_{\rm unit}/q}}\right)^2\sum_{\bm{K},\bm{K'}}e^{-i(\bm{K}-\bm{K}')\cdot\bm{d}}\langle\bar{\gamma}_{\bm{K},a}\gamma_{\bm{K}',a}\rangle\notag\\
    &=\frac{q}{N_{\rm unit}}\sum_{\bm{K}}\langle\bar{\gamma}_{\bm{K},a}\gamma_{\bm{K},a}\rangle\notag\\
    \Leftrightarrow&\sum_{\bm{K}}\langle\bar{\gamma}_{\bm{K},a}\gamma_{\bm{K},a}\rangle=\frac{N_{\rm unit}}{q^2},\label{constraint}
\end{align}
where we have used $\langle\bar{\gamma}_{\bm{K},a}\gamma_{\bm{K}',a}\rangle\propto\delta_{\bm{K}\bm{K}'}$.
In the actual numerical calculations, we check this constraint after the self-consistent procedure.

\subsection{Non-Hermitian Chern number}
In the conventional composite-fermion approach, the FQHE with $\nu=1/q$ is interpreted as the IQHE of the composite fermions \cite{jain2007composite}.
Thus, the FCI with $\nu=1/q$ is expected to correspond to the Chern insulator with $C=1$.
In our formalism, however, $H^{ab}_{\bm{K}}$ is a non-Hermitian matrix, and the Chern number should be defined in a non-Hermitian manner.
Note that the Hermitian conjugate of a ket state (right eigenvector) is not always a bra state (left eigenvector) under the non-Hermiticity.

We extend the Fukui-Hatsugai formula \cite{Fukui-Hatsugai-Suzuki-05} to the non-Hermitian Bloch Hamiltonian:
\begin{align}
    &C=\frac{1}{2\pi }\sum_{\bm{K}\in \mathrm{MBZ}}F_{12}(\bm{K}),\label{nhchern}\\
    &F_{12}(\bm{K})=\mathrm{Arg}~ U_1(\bm{K})U_2(\bm{k}+\hat{1})U^{-1}_1(\bm{K}+\hat{2})U^{-1}_2(\bm{K}),\\
    &U_i(\bm{K})=\langle   L,\bm{K}|R,\bm{K}+\hat{i}    \rangle,
\end{align}
where MBZ denotes the magnetic Brillouin zone of composite fermions, $\hat{i}$ is a vector in the direction $i$ with the magnitude $2\pi/L_i$, and $-\pi<\mathrm{Arg} ~z\leq \pi$. $F_{12}$ describes the discretized Berry curvature.
The left and right eigenvectors, $\bra{L,\bm{K}}$ and $\ket{R,\bm{K}}$, satisfy the followings:
\begin{align}
    &\hat{H}_{\bm{K}}\ket{R,\bm{K}}=E_{\bm{K}}\ket{R,\bm{K}},\\
    &\bra{L,\bm{K}}\hat{H}_{\bm{K}}=E_{\bm{K}}\bra{L,\bm{K}},\\
    &\langle L,\bm{K}\ket{R,\bm{K}}=1,~\langle R,\bm{K}\ket{R,\bm{K}}=1,
\end{align}
where $E_{\bm{K}}$ is the eigenenergy of the band of interest.
In the infinite-volume limit, this formula corresponds to the non-Hermitian Chern number \cite{ksus}:
\begin{align}
    &C=\frac{1}{2\pi }\int_{MBZ}d^2K~\left[\partial_xA^{LR}_y(\bm{K})-\partial_yA^{LR}_x(\bm{K})\right],\\
    &A^{LR}_i(\bm{K})=-i\bra{L,\bm{K}}\partial_i\ket{R,\bm{K}}.
\end{align}

\subsection{Numerical results}

In this subsection, we perform the non-Hermitian Hartree-Fock mean-field calculation for two models that are defined on a square lattice and include two orbital degrees of freedom within the unit cell. In the absence of Hermiticity, there are a lot of self-consistent solutions even under the constraint (\ref{constraint}), and many of the self-consistent spectra are complex.
Although we do not have any logical reasons, following the steps below tends to reduce the real part of the ground state energy within our search range.
Even more surprisingly, such spectra are mostly located on the real axis.
We first calculate the self-consistent solution of $\langle\bar{\gamma}_{\bm{K},a}\gamma_{\bm{K},b}\rangle$ without the last constant term of Eq. (\ref{hamexpression}).
In our numerical observation, the obtained spectrum is almost on the real axis.
Then, using the obtained $\langle\bar{\gamma}_{\bm{K},a}\gamma_{\bm{K},b}\rangle$ as the initial value for the self-consistent loop, we perform the true mean-field calculation.

\begin{figure}[]
\begin{center}
 \includegraphics[width=8.5cm,angle=0,clip]{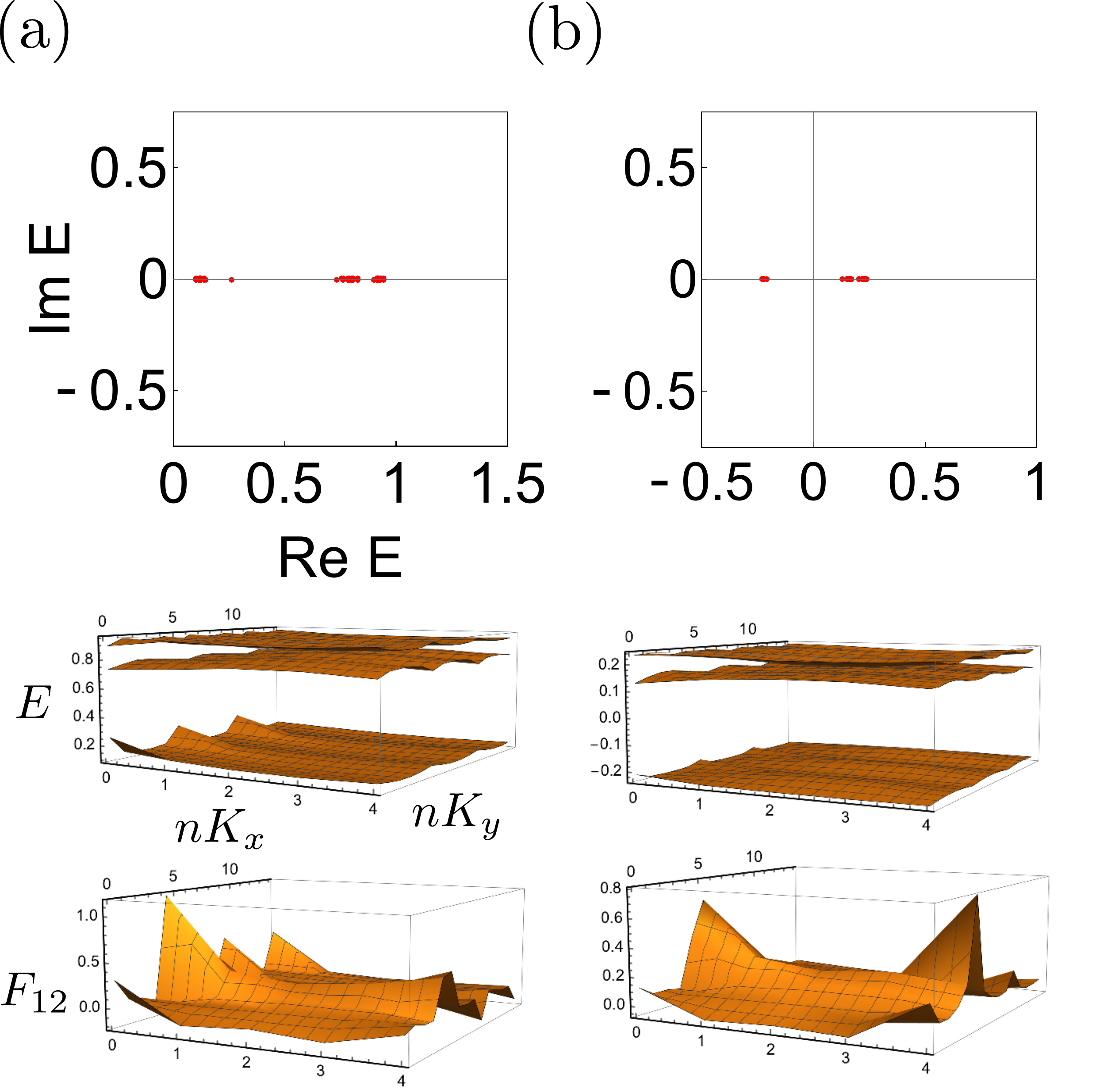}
 \caption{Energy spectra, dispersions, and discretized Berry curvature in non-Hermitian Hartree-Fock calculation for (a) checkerboard-lattice and (b) QWZ models. Discretized momenta are labeled by $(nK_x,nK_y)$.}
 \label{fig4}
\end{center}
\end{figure}

\begin{figure}[]
\begin{center}
 \includegraphics[width=8cm,angle=0,clip]{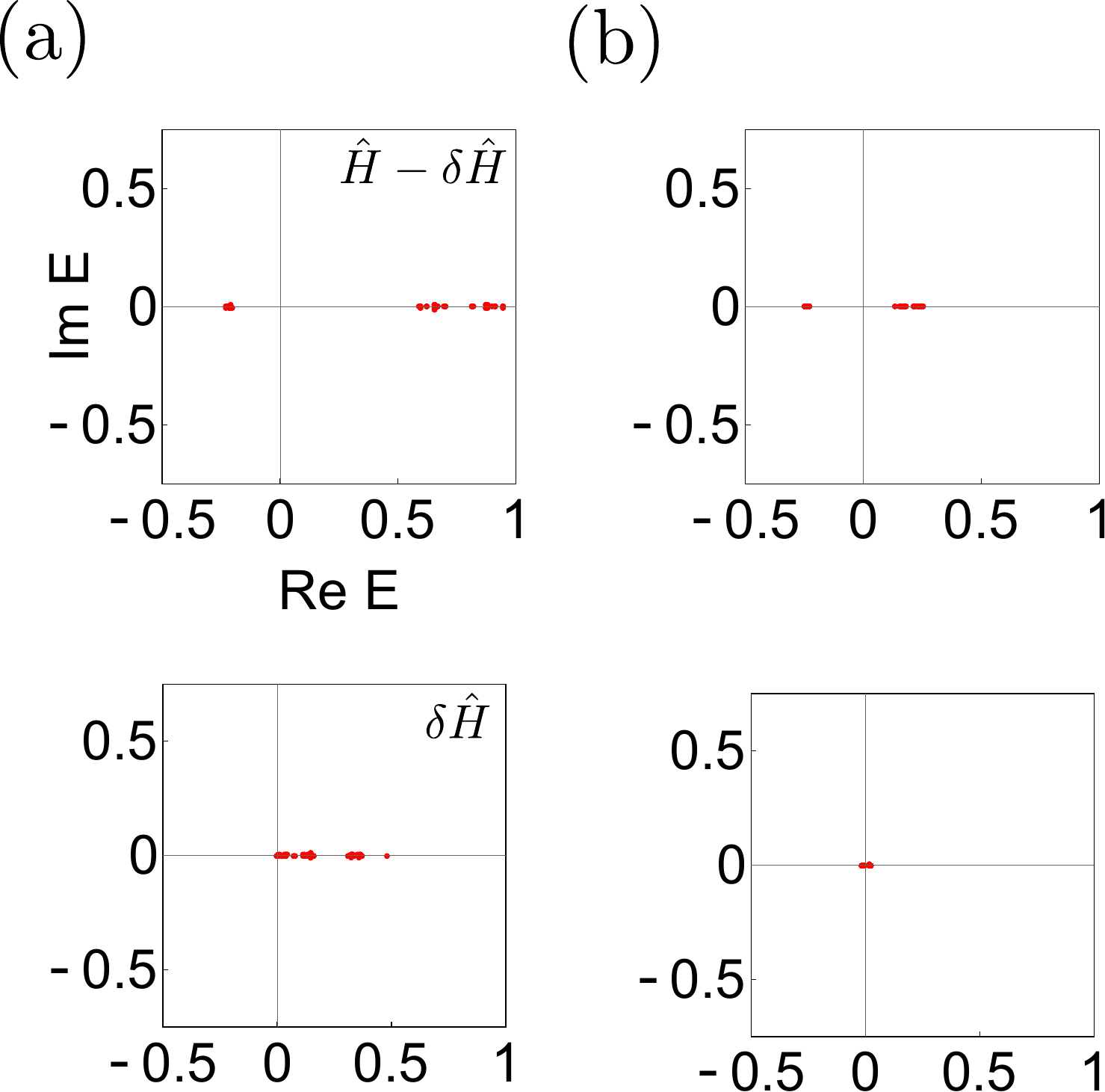}
 \caption{Spectra of $\hat{H}_{\bm{K}}-\delta\hat{H}_{\bm{K}}$ and $\delta\hat{H}_{\bm{K}}$ for (a) checkerboard-lattice and (b) QWZ models. }
 \label{fig5}
\end{center}
\end{figure}

For the concrete calculations, we consider the checkerboard-lattice model \cite{Neupert-Santos-Chamon-Mudry-11}, which has already been investigated in Sec.\ref{biorthogonalsection}, and the Qi-Wu-Zhang (QWZ) model ~\cite{Qi-Wu-Zhang-06}.
See the Appendix for details of the models.
As parameters of the lattice vortex function, we use $(\alpha_1,\alpha_2)=(1,i)$ for both models.
We set $\{\tilde{\bm{r}}_1,\tilde{\bm{r}}_2\}=\{(0.5,0),(0,0.5)\}$ for the checkerboard-lattice model and $=\{(0,0),(0,0)\}$ for the QWZ model, both of which are identical to the real sublattice positions, $\{\bm{r}_1,\bm{r}_2\}$.
As the repulsive interaction, we use the nearest-neighbor interaction for the checkerboard-lattice model and the on-site interaction between the sublattice 1 and 2 for the QWZ model. The strength of the interaction is set to unity.
The number of unit cells is $15\times15$, and the filling factor $\nu$ is $1/3$. Then, the number of particles is $75$.
In Fig.\ref{fig4}, we plot the energy spectrum, dispersion, and the lowest-band discretized Berry curvature.
In both cases, the non-Hermitian Chern number (\ref{nhchern}) of the lowest band becomes unity, and the constraint (\ref{constraint}) is satisfied. 
These behaviors are consistent with the belief that the FCI ground state can be regarded as a composite-fermion Chern insulator.
As mentioned above, the energy spectra are almost on the real axis.
While all the eigenenergies are positive for the checkerboard-lattice model, the lowest-band energies are negative for the QWZ model.
Thus, the ground-state energy for the QWZ model becomes negative, though the original Hamiltonian before the approximations is positive definite.
In this sense, the result for the QWZ model is pathological.
The difference between the two models comes from the amount of the constant term:
\begin{align}
    \sum_{\bm{K}',c} V^{a,c,c,b}_{\bm{K},\bm{K}',\bm{K}',\bm{K}}=:\delta H^{ab}_{\bm{K}}.
\end{align}
In Fig.\ref{fig5}, we show the spectra of $\delta\hat{H}_{\bm{K}}$ and the self-consistent spectra of $\hat{H}_{\bm{K}}-\delta\hat{H}_{\bm{K}}$.
In both cases, $\hat{H}_{\bm{K}}-\delta\hat{H}_{\bm{K}}$ has negative eigenvalues.
While the matrix $\delta\hat{H}_{\bm{K}}$ has a sufficiently large magnitude to lift those negative eigenvalues to positive ones in the checkerboard-lattice model, it does not in the QWZ model.

\section{Discussion}
\begin{figure}[]
\begin{center}
 \includegraphics[width=8.5cm,angle=0,clip]{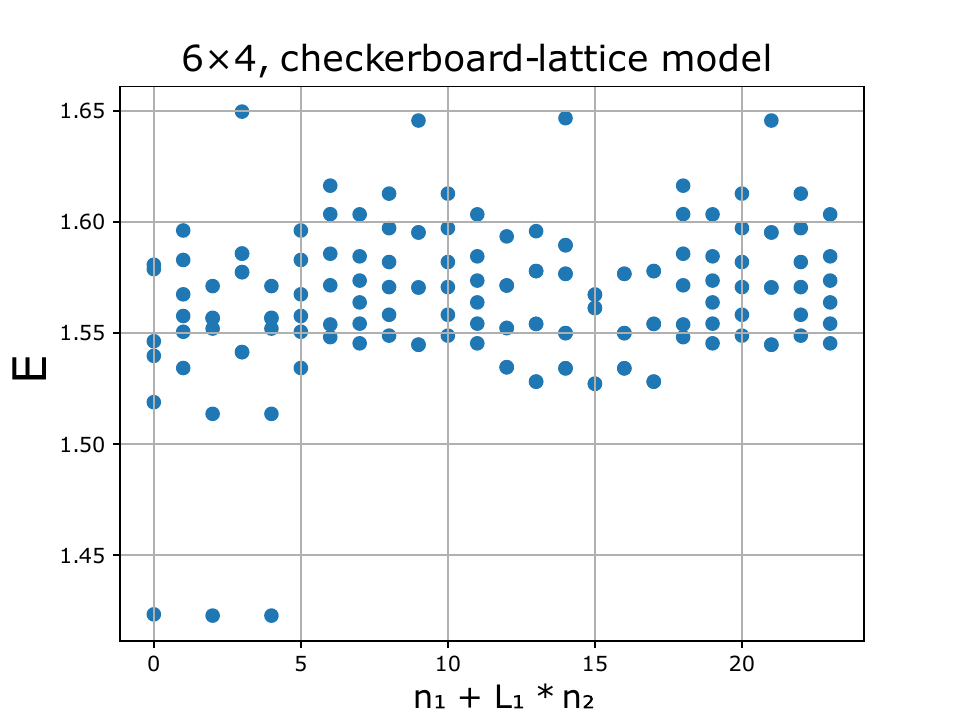}
 \caption{Exact diagonalization for checkerboard-lattice model with $L_1\times L_2=6\times4$ unit cells. Small eigenvalues are shown. Total momentum sectors are labeled by $(n_1,n_2)$. }
 \label{fig6}
\end{center}
\end{figure}

Before closing the paper, we discuss several points.
At this stage, it is unclear whether the sign of the ground state energy of our composite fermions is directly related to the validity of the approximation.
If it describes the FCI stability, one can say that the checkerboard-lattice model is a good FCI candidate, and the QWZ model is a bad one.
This observation coincides with the studies of ideal conditions of Chern bands \cite{ledwith2023vortexability, Roy-geometry-14, Jackson-Moller-Roy-15}, including our previous estimation of lattice vortexability \cite{okuma2024constructing}.
This indicates that the QWZ model seems more likely to show non-FCI ground states, at least in comparison with other models such as the checkerboard-lattice model.

Note that even if one were to believe that the above sign is useful, the quantitative nature of this approximation would hardly be described as good. In Fig.\ref{fig6}, we plot the exact-diagonalization spectrum of the checkerboard-lattice model with $6\times4$ unit cells under the projection.
If we convert the results of our approximation to the same size, we find that the ground-state energy is underestimated, and the energy gap is overestimated. 
The modification of quantitative precision is a major future work.
It might be helpful to compare with the more historically established Hamiltonian theory \cite{murthy2003hamiltonian,murthy2012hamiltonian,hu2024hyperdeterminants}.
 
One can also consider the composite bosonization on our biorthogonal basis.
Under a similar procedure, the effective Hamiltonian is mapped to a hardcore bosonic model or a spin-one-half four-body spin model.
Ideally, the FCI ground state should correspond to a Bose-Einstein condensate or a canted ferromagnetism.
Although approximating the troublesome four-body interaction in a spin model is more unconventional than in the case of composite fermions, it is also possible that the accuracy of the approximation could improve.

A generalization for $\mathbb{Z}_2$ topological insulators is another interesting direction.
In our previous work \cite{okuma2024constructing}, we pointed out that the $\mathbb{Z}_2$ index theorem under the fermionic time-reversal symmetry enables us to consider the von Neumann lattice for the topological insulators.
By defining a biorthogonal lattice for such a case, it may be possible to consider composite fermionization for a fractional topological insulator \cite{levin2009fractional,stern2016fractional}.

\acknowledgements
I thank Avedis Neehus for the e-mail discussions on finite systems.
This work was supported by JSPS KAKENHI Grant No.~JP20K14373 and No.~JP23K03243.
%\textcolor{red}{Z for specific momentum is always optimized if the band number is one. weight no kotonaru baai no Z wo kangaeru. Appendix}
\\
\appendix
\section{Models}
In this section, we write down the Bloch Hamiltonians for the Chern insulators used in the main text.
\subsection{Checkerboard-lattice model}
This model was studied in terms of the fermionic FCI \cite{Neupert-Santos-Chamon-Mudry-11}. 
The Bloch Hamiltonian matrix is given by
\begin{align}
    H(\bm{k})=
    \begin{pmatrix}
    2t_2(\cos k_1-\cos k_2)&t_1f^*(\bm{k})\\
    t_1f(\bm{k})&-2t_2(\cos k_1-\cos k_2)\\
    \end{pmatrix},
\end{align}
where
\begin{align}
    &f(\bm{k})=e^{-i\pi/4}\left[1+e^{i(k_2-k_1)}\right]+e^{i\pi/4}\left[e^{-ik_1}+e^{ik_2}\right].
\end{align}
In numerical calculations, we set $t_1=1,t_2=\sqrt{2}/2$. For this parameter, the band structure becomes almost flat.

\subsection{Qi-Wu-Zhang (QWZ) model}
This model was proposed in Ref.~\cite{Qi-Wu-Zhang-06} and is also referred to as the Wilson-Dirac model. 
The Bloch Hamiltonian matrix is given by
\begin{align}
    &H(\bm{k})=\sin k_1 \sigma_x+\sin k_2 \sigma_y+(m-\cos k_1-\cos k_2)\sigma_z, \label{eq:QWZ_Bloch}
\end{align}
where $\sigma_{x,y,z}$ are the Pauli matrices. 
In numerical calculations, we set $m=1$.

\bibliography{FCI}
\end{document}